\documentclass[twoside,10pt,a4paper]{newFNLstyle}
\usepackage{graphics}
\usepackage{cite}

\begin{document}

\volnumpagesyear{0}{0}{000--000}{2001}
\dates{received date}{revised date}{accepted date}

\title{FLUCTUATIONS AND NOISE IN TIME-RESOLVED LIGHT SCATTERING EXPERIMENTS: MEASURING TEMPORALLY HETEROGENEOUS DYNAMICS}

\authorsone{A. Duri, P. Ballesta, L. Cipelletti}
\affiliationone{GDPC UMR 5581 Universit\'{e} Montpellier II et CNRS, P. E. Bataillon, Montpellier, France;}
\mailingone{lucacip@gdpc.univ-montp2.fr}

\authorstwo{H. Bissig, V. Trappe}
\affiliationtwo{Department of Physics, University of Fribourg, Perolles,Fribourg, Switzerland}


\maketitle

\markboth{Fluctuations and noise in time-resolved light scattering}{A. Duri, P. Ballesta, L. Cipelletti, H. Bissig and V. Trappe}

\pagestyle{myheadings}


\begin{abstract}
We use Time Resolved Correlation (TRC), a recently introduced
light scattering method, to study the dynamics of a variety of
jammed, or glassy, soft materials. The output of a TRC experiment
is $c_I(t,\tau)$, the time series of the degree of correlation
between the speckle patterns generated by the light scattered at
time $t$ and $t+\tau$. We characterize the fluctuations of $c_I$
by calculating their Probability Density Function, their variance
as a function of the lag $\tau$, and their time autocorrelation
function. The comparison between these quantities for a Brownian
sample and for jammed materials indicate unambiguously that the
slow dynamics measured in soft glasses is temporally
heterogeneous. The analogies with recent experimental, numerical
and theoretical work on temporal heterogeneity in the glassy
dynamics are briefly discussed.
\end{abstract}

\section{Introduction}
\label{sect:intro}  

Dynamic light scattering (DLS), also known as photon correlation
spectroscopy, is a popular technique that provides information on
the dynamics of soft condensed matter systems \cite{Berne1976}. It
is widely used in physics, chemistry, and biology to investigate
systems as diverse as colloidal particle suspensions, emulsions,
polymer or protein solutions, surfactant phases, and gels. In the
past years, new methods and techniques have been developed that
have greatly extended the range of applicability of dynamic light
scattering. The advent of coherent, high brilliance X-ray sources
has allowed photon correlation spectroscopy experiments to be
performed using X-ray radiation \cite{DierkerPRL1995}, instead of
the laser beam used for DLS. Since in a scattering experiment the
wavelength of the radiation sets the length scale over which the
dynamics is probed, X-ray photon correlation spectroscopy (XPCS)
is a significant extension of DLS. The interpretation of both DLS
and XPCS data is based on the assumption that the scattering is
sufficiently weak, so that virtually all photons collected by the
detector are singly scattered and the distortions induced by
multiple scattering can be neglected. Diffusing Wave Spectroscopy
(DWS) \cite{DWSGeneral}, on the contrary, allows one to study very
turbid samples, for which multiple scattering dominates, thus
extending the range of samples to which dynamic scattering methods
can be applied.

Quite generally, in a scattering experiment the information on the
dynamics is extracted from the time fluctuations of the scattered
intensity, $I(t)$. The intensity is usually measured by a point
detector, such as a phototube; its fluctuations are quantified by
calculating the time autocorrelation function of $I(t)$, defined
by $G_2(\tau) = \langle I(t)I(t+\tau)\rangle_t$, where
$\langle...\rangle_t$ denotes an average over
time\footnote{Although in principle information on the dynamics
may also be extracted from the spectral properties of the
scattered light, the analysis in the temporal domain has become
the standard procedure.}. The time average is necessary to obtain
sufficiently good statistics: typically, data have to be collected
over a period as long as $10000$ times the relaxation time of
$G_2(\tau)$. Because of the extended time averaging, traditional
scattering methods can only provide information on the $average$
dynamics of a system, with no time resolution. The lack of time
resolution, together with the need of collecting data over a long
period, makes traditional scattering techniques difficult to apply
to glassy
---or jammed \cite{LiuNature1998}--- systems, for which there is
currently a great interest (for a review, see for example
Ref.~\citenum{LucaCOCIS2002}). In fact, for these systems the
relaxation time can be as long as several thousand seconds, thus
making time averaging practically impossible. Moreover, their
dynamics is often non-stationary; therefore, time-averaged
correlation functions do not allow the evolution of the dynamics
to be followed precisely. The introduction of the so-called
multispeckle technique has addressed these issues. In a
multispeckle experiment, the point detector is replaced by the 2-D
array of pixels of a CCD camera
\cite{WongRSI1993,KirschJChemPhys1996,LucaRSI1999}. Each CCD pixel
acts as an independent detector, for which an intensity time
autocorrelation function is calculated. In multispeckle
experiments, the time average that is usually performed in
traditional light scattering is replaced by an average over the
CCD detector: $G_2(t,\tau) = \langle I_p(t)I_p(t+\tau)\rangle_p$,
where the subscript $p$ refers to the CCD pixels. In most works,
$G_2(t,\tau)$ is also averaged over a short time window
\cite{LucaRSI1999} or over several repetitions of the experiment
\cite{ViasnoffPRL2002}, in order to reduce the measurement
``noise''.

Recently, we have proposed a novel scattering method, termed Time
Resolved Correlation (TRC), where valuable physical information on
the dynamics are extracted precisely from this ``noise'', i.e.
from the time fluctuations of $G_2(t,\tau)$ \cite{LucaJPCM2003}.
In particular, we have shown that, contrary to traditional
scattering methods, TRC allows temporally heterogeneous dynamics
to be discriminated from homogeneous dynamics\footnote{As it is
common practice in the literature on glassy systems, in this paper
we use the adjective ``heterogeneous'' to indicate the presence of
fluctuations in the dynamics other than the inevitable noise
associated with the measuring process. Note that, with this
definition, both stationary and non-stationary dynamical processes
may be temporally heterogeneous.}. Because there is growing
evidence that dynamical heterogeneity is a key feature of the slow
dynamics of glassy systems, TRC appears as a very promising tool
for investigating the slow relaxation of these systems. In this
paper, we review the principles of TRC and we propose some ways of
analyzing TRC data. The paper is organized as follows:
Sect.~\ref{TRCSec} briefly reviews the basic features of a
scattering experiments and the way TRC data are collected. In
Sect.~\ref{Homo} we develop the statistical analysis used to
characterize the fluctuations of the TRC signal and we apply it to
data collected for systems whose dynamics is homogeneous. In the
following Section, the same tools are used for systems that
exhibit temporal heterogeneities. Section~\ref{Comp} concludes the
paper by summarizing the main results and briefly compares them to
other experimental and numerical works on systems whose dynamics
is temporally heterogeneous.

\section{Time Resolved Correlation}
\label{TRCSec}

The TRC method can be applied both to the single scattering
geometry and to DWS experiments. In all cases the multispeckle
detection is adopted: a CCD camera is used to take a time series
of images of the speckle pattern, the characteristic grainy
picture generated by the interference of the photons scattered by
the sample.

\begin{figure}[htbp]
\centering{\resizebox{12.5cm}{!}{\includegraphics{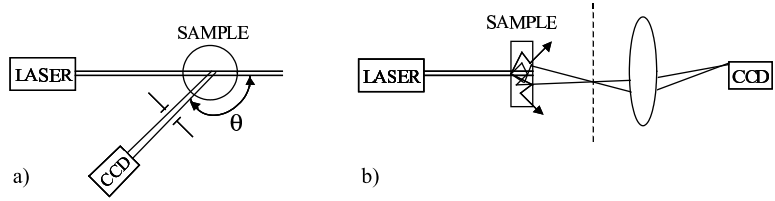}}}
\caption{\label{apparatus}
Schematic representation of a) a single scattering apparatus, b) a
multiple scattering (DWS) setup in the transmission geometry. In
a), $\theta$ is the scattering angle, selected by the diaphragm in
front of the CCD . The lens in b) conjugates the plane indicated
by the dashed line with the CCD chip. In both cases, the CCD
camera is used to record the speckle pattern generated by the
interference of the scattered photons.}
\end{figure}

Typical scattering setups are sketched in Fig.~\ref{apparatus},
for single (a) and multiple (b) scattering measurements; XPCS
instruments are similar, the laser source being replaced by
synchrotron radiation. A detailed discussion on how the speckle
pattern fluctuates because of the sample dynamics is beyond the
scope of this paper and can be found in
Ref.~\citenum{Berne1976,DWSGeneral}. For our purposes, we just
note that any change in the position of the scatterers modifies
the relative phase of the scattered photons and thus the intensity
pattern of the speckles generated by their interference. Using the
notation of Ref.~\citenum{LucaJPCM2003}, we quantify the change in
the speckle pattern by introducing $c_I(t,\tau)$, the normalized
intensity correlation between speckle images at time $t$ and
$t+\tau$:
\begin{equation}
\label{eq:cI} c_I(t,\tau) = \frac{G_2(t,\tau)}{\langle
I_p(t)\rangle_p \langle I_p(t+\tau)\rangle_p} -1 = \frac{\langle
I_p(t)I_p(t+\tau)\rangle_p}{\langle I_p(t)\rangle_p \langle
I_p(t+\tau)\rangle_p} - 1\, ,
\end{equation}
where the average is over CCD pixels. The normalization in
Eq.~(\ref{eq:cI}) guarantees that $c_I(t,\tau)$ is not affected by
any change in the incoming beam intensity. Because of the
statistical properties of the speckle pattern, $0 \leq c_I(t,\tau)
\leq \beta$, where $\beta \leq 1$ is a prefactor that depends on
the setup geometry \cite{Berne1976,Goodman1975}.

\begin{figure}[htbp]
\centering{\resizebox{12.5cm}{!}{\includegraphics{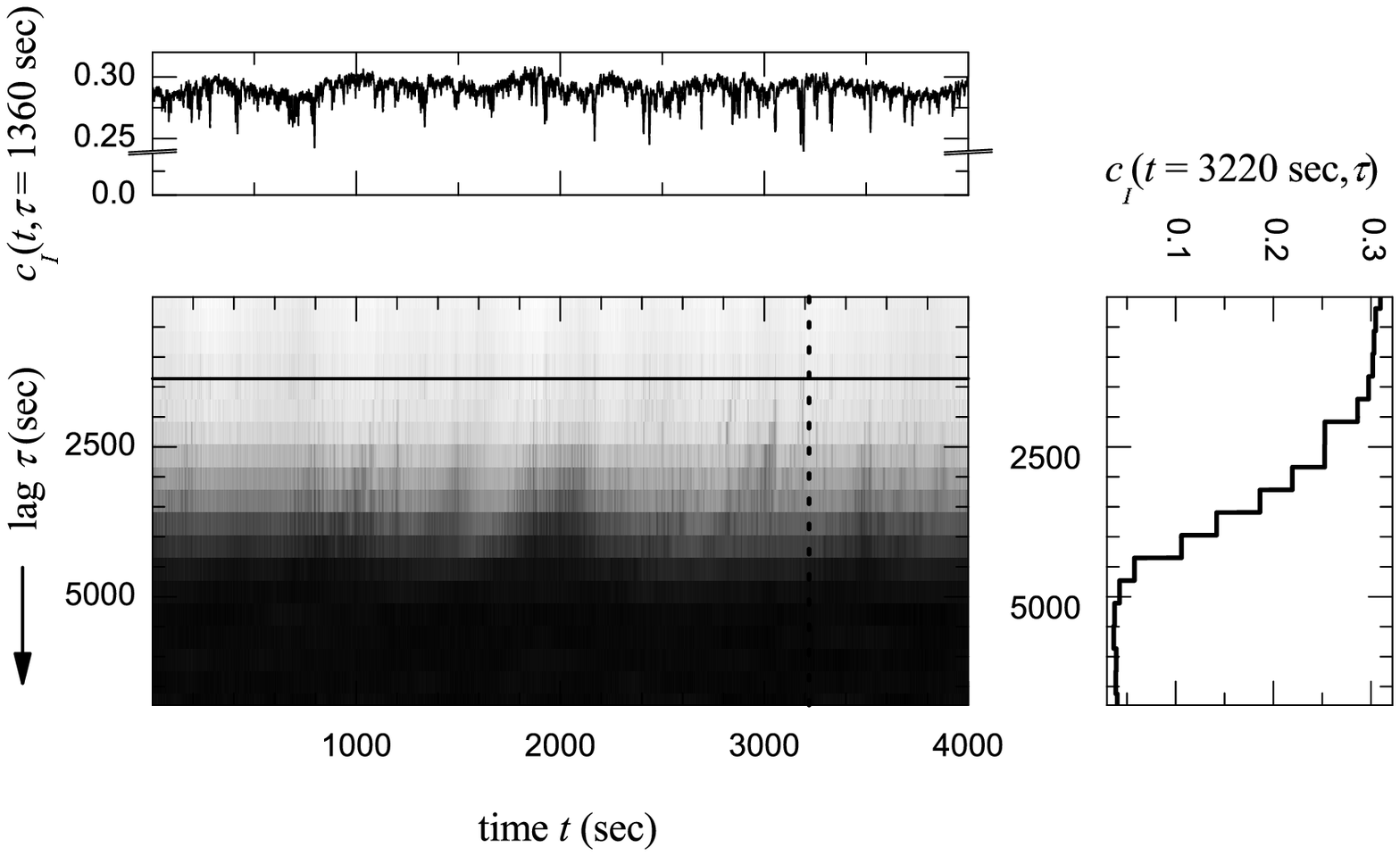}}}
\caption{\label{cI2D}
Bottom left panel: 2-D grey scale representation of $c_I(t,\tau)$,
the degree of correlation between speckle patterns recorded at
time $t$ and $t+\tau$. White is the highest correlation, black the
lowest. The sample is a concentrated colloidal paste. Right panel:
two-time intensity autocorrelation function obtained by cutting
$c_I$ along the vertical dashed line. Top panel: temporal
fluctuations of $c_I$ at a fixed lag time, obtained from a cut
along the horizontal solid line. Note the sharp drops of
correlation, indicative of sudden, temporally localized
rearrangements.}
\end{figure}

Figure~\ref{cI2D} shows a 2-D grey scale representation of $c_I$
as a function of both time $t$ and time lag $\tau$ (data were
taken in the DWS geometry for a colloidal paste
\cite{BallestaJapan2004}). White corresponds to the highest degree
of correlation, $c_I = \beta$, while black corresponds to
uncorrelated pairs of speckle patterns, $c_I = 0$. As general
trend, for any given $t$ $c_I$ decreases with increasing time lag
$\tau$, since the longer the lag, the greater the change of the
system configuration and thus that of the speckle images. This is
exemplified by the right panel that shows
$c_I(t=3220~\rm{sec},\tau)$, obtained by cutting the 2-D data
along the vertical dashed line. Note that the time average of
$c_I(t,\tau)$ yields the normalized intensity autocorrelation
function, $g_2(\tau)-1$, usually measured in light scattering
experiments\footnote{In traditional light scattering measurements,
the average is exclusively over time when using a point detector,
over both pixels and time when using a CCD camera.}. The time
fluctuations of $c_I$ measured at a constant lag
$\tau=1360~\rm{sec}$, obtained from a cut of the 2-D data along
the horizontal solid line, are shown in the top panel. Sharp drops
of correlation that depart from the typical fluctuations of the
signal are clearly visible. These drops suggest that occasionally
the change in the speckle pattern, and thus in sample
configuration, is much larger than average: they are therefore
indicative of a temporally heterogeneous dynamics. In the rest of
the paper, we will focus on the statistical analysis of the
fluctuations of $c_I(t,\tau)$ at fixed lag $\tau$, first for
samples that exhibit homogeneous dynamics, and then for systems
whose dynamics is temporally heterogeneous.

\section{Statistical Analysis of TRC Data for temporally homogeneous dynamics}
\label{Homo}

In order to better appreciate the signature of heterogeneous
dynamics in TRC data, we first characterize the statistics of the
fluctuations of $c_I(t,\tau)$ in the case of temporally
homogeneous dynamics. The data presented in this Section are taken
in the single scattering geometry, at a scattering angle $\theta =
90$~deg\footnote{Similar results are obtained in DWS}. The sample
is a diluted suspensions of particles undergoing Brownian motion:
polystyrene spheres of radius $r = 530~\rm{nm}$ were suspended at
a volume fraction $\phi = 3.7 \times 10^{-5}$ in almost pure
glycerol (approximately 99.9/0.001 w/w glycerol/water). The sample
temperature was controlled and fixed at $T = 15 \pm
0.1~\rm{^o}\rm{C}$.

   \begin{figure}
\centering{\resizebox{12.5cm}{!}{\includegraphics{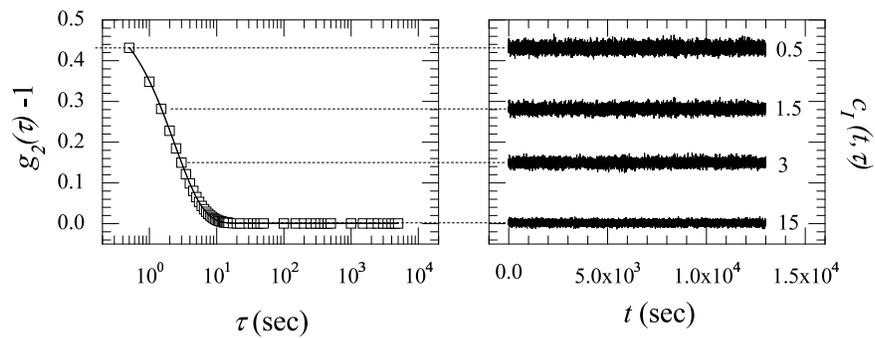}}}
\caption{\label{cIpoly}
For a dilute suspension of Brownian particles are shown: left
panel: the intensity autocorrelation function $g_2(\tau)-1$; right
panel: $c_I(t,\tau)$ as a function of $t$ for four time lags
$\tau$. The correlation function shown in the left panel was
obtained by time-averaging the TRC data plotted in the right panel
(for clarity, not all the time lags are shown in the right
panel).}
   \end{figure}

Figure~\ref{cIpoly} a) shows the time-averaged intensity
autocorrelation function $g_2(\tau)-1$ obtained from the TRC data
for the Brownian particles. The data are well fitted by an
exponential decay $\beta \exp(-\tau/\tau_s)$ (line in the figure),
as expected for a dilute suspension of monodisperse spheres
\cite{Berne1976}, with $\tau_s$ = 2.35 sec. The right panel shows the
corresponding $c_I(t,\tau)$ as a function of time $t$ for four
time lags, ranging from $\tau/\tau_s = 0.21$ to $\tau/\tau_s =
6.38$. For Brownian particles, we expect the dynamics to be
temporally homogeneous. Accordingly, the loss of correlation
between pairs of speckle images separated by a given time lag
$\tau$ should not depend on the time the first image of the pair
is taken. In other words, for a fixed lag $\tau$, $c_I(t,\tau)$
should be constant. This is indeed the case, except for the small
fluctuations visible in in the right panel, which we attribute
to the CCD noise and the finite number of pixels, and therefore
independent speckles, over which $c_I$ is averaged. In the
following we shall refer to both sources of noise as to the
``measurement noise''.

  \begin{figure}
\centering{\resizebox{12.5cm}{!}{\includegraphics{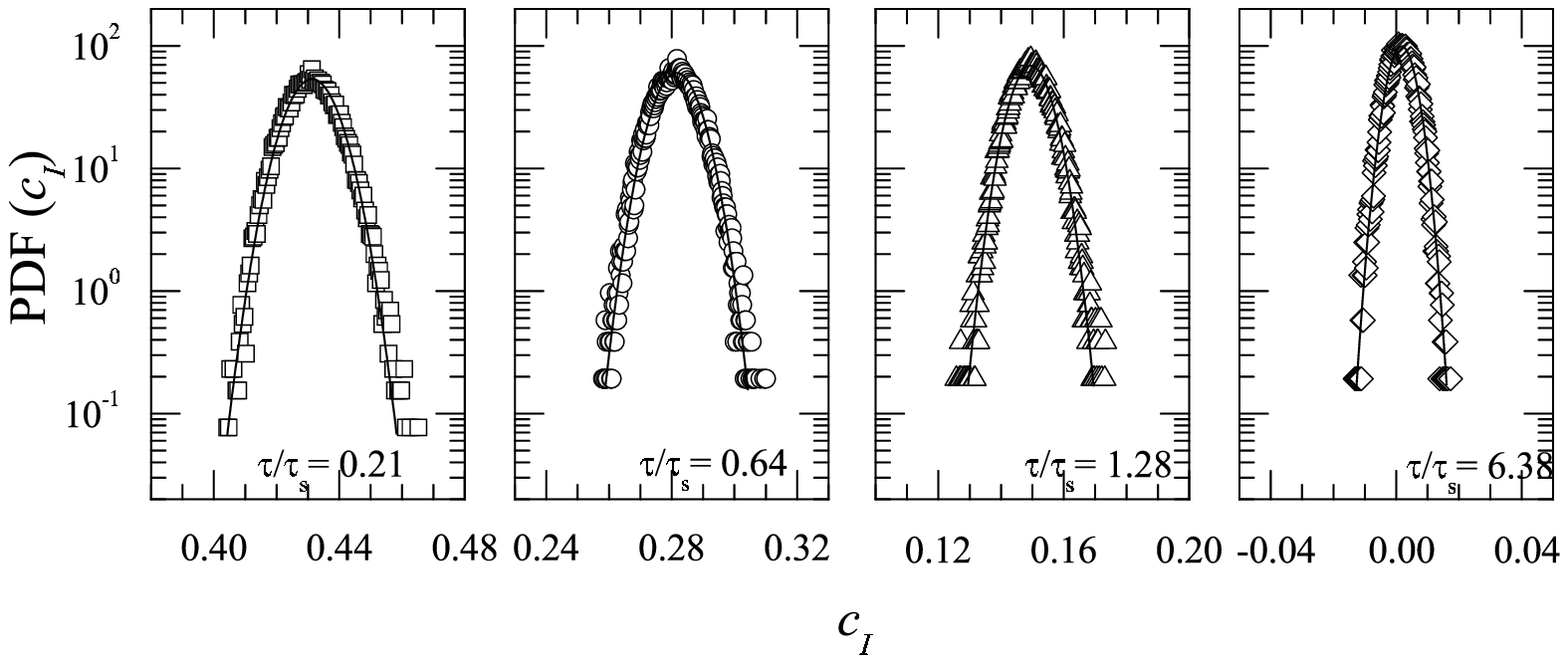}}}
\caption{\label{PDFpoly}
Probability Density Function of the temporal fluctuations of
$c_I(t,\tau)$, for the data shown in Fig.~\ref{cIpoly} b)
(symbols: from left to right, $\tau$ = 0.5, 1.5, 3, and 15 sec).
The lines are Gaussian fits to the data. Note that width oh the
PDF's ---and thus the standard deviation of the fluctuations---
decreases at larger lags, together with the mean of $c_I$.}
   \end{figure}

\begin{figure}
\centering{\resizebox{12.5cm}{!}{\includegraphics{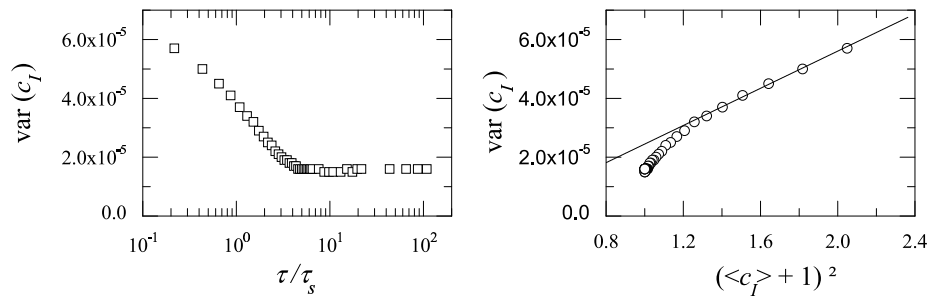}}}
\caption{ \label{VarVsTau}
Left: lag dependence of the variance of the temporal fluctuations
of $c_I$ for a  suspensions of Brownian particles. Right: same
data plotted parametrically $vs$ the squared mean of the TRC
signal, $(\langle c_I \rangle_t +1)^2$. Except for the lowest
values of $(\langle c_I \rangle_t +1)^2$, the data follow a linear
behavior, as shown by the straight line.}
   \end{figure}

\begin{figure}
\centering{\resizebox{12.5cm}{!}{\includegraphics{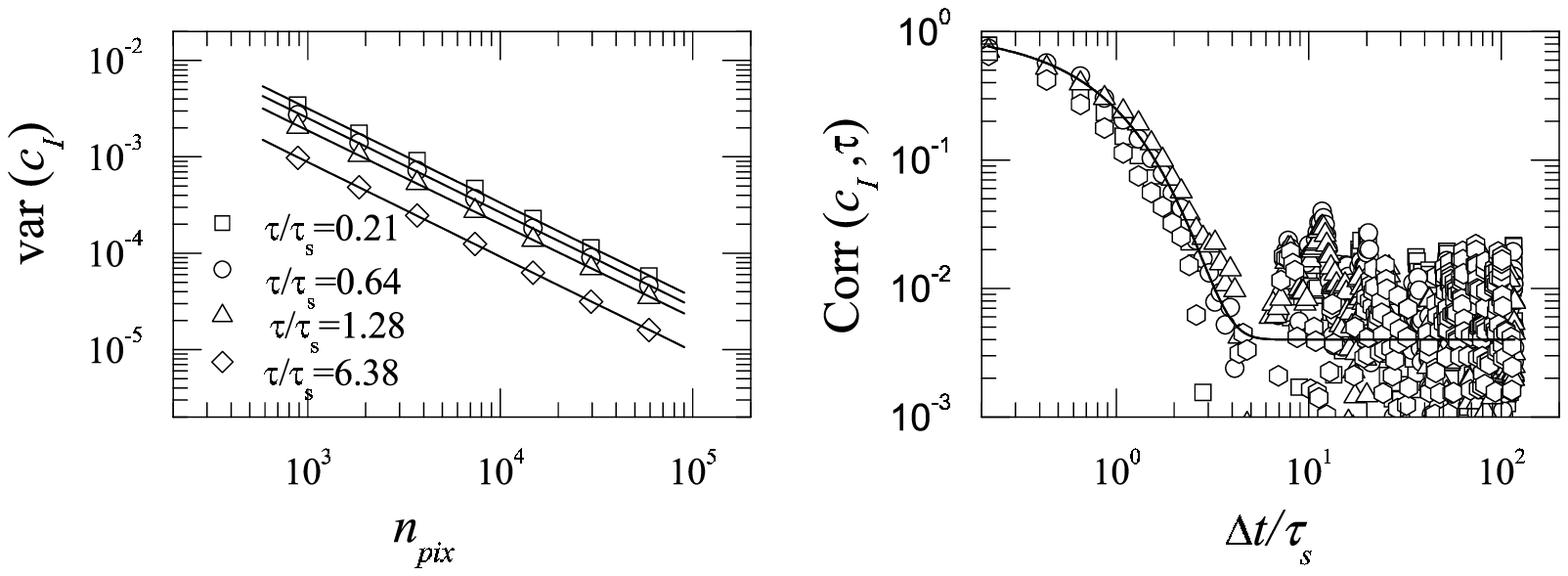}}}
\caption{ \label{VarVsnpix}
For a suspension of Brownian particles are shown: Left panel:
double logarithmic plot of the variance of the temporal
fluctuations of $c_I$ , as a function of the number of processed
pixels, and for various time lags. For all $\tau$, the data are
well fitted by a power law decay with exponent $-0.98 \pm 0.01$,
very close to what is expected for the measurement noise because of
the Central Limit Theorem. Right panel: time autocorrelation
function of the fluctuations of $c_I$, for various lags (symbols).
The line is an exponential fit to the data for $\tau = 1.5$ sec
(the fit for the other values of $\tau$ are omitted for clarity).
The correlation in the fluctuations stems from the finite number
of pixels over which $c_I$ is averaged, as discussed in the text.}
   \end{figure}

In order to verify this hypothesis, we analyze in some detail the
temporal fluctuations of $c_I$. Figure~\ref{PDFpoly} shows the
Probability Density Function (PDF) of $c_I$ for the same time lags
as in Fig.~\ref{cIpoly}, right panel. At all lags, the data are well fitted
by a Gaussian distribution, as shown by the lines. This behavior
is consistent with the hypothesis that the fluctuations are due to
the measurement noise. Note that the width of the distributions
decreases for increasing lag $\tau$, together with the mean value
$g_2(\tau)-1 = \langle c_I \rangle_t$.To better investigate the lag
dependence of the amplitude of the fluctuations, we plot in
the left panel of Fig.~\ref{VarVsTau} the variance of $c_I$, ${\rm
var}(c_I)$, as a function of $\tau$ normalized by the mean
relaxation time $\tau_s$. At the smaller lags, the decay of the
amplitude of the fluctuations closely follows that of
$g_2(\tau)-1$, whereas for $\tau > \tau_s$ ${\rm var}(c_I)$
saturates to a plateau. This behavior suggests the following form
for the variance of $c_I$: ${\rm var}(c_I) = A + B(\langle c_I
\rangle_t +1)^2$, where the second term of the r.h.s. represents a
noise contribution proportional to $G_2(t,\tau) / \langle
I_p(t)\rangle_p \langle I_p(t+\tau)\rangle_p$, the signal that is
measured in a TRC experiment. The first term is the noise of $c_I$
when all correlation between speckle images is lost, i.e. the fluctuations
of the base line for $\tau \rightarrow \infty$. As it will be
discussed in more detail in reference to Fig.~\ref{VarVsnpix},
this contribution is mainly due to the finite number of pixels.
The right panel of Fig.~\ref{VarVsTau} shows ${\rm var}(c_I)$ $vs$
$(\langle c_I \rangle_t +1)^2$ (symbols). The data are in good
agreement with the linear behavior ${\rm var}(c_I) = A + B(\langle
c_I \rangle_t +1)^2$ introduced above, except for the downturn
visible at the smallest values of $\langle c_I \rangle_t$ (the
straight line is a linear fit for $(\langle c_I \rangle_t +1)^2 >
1.3$). The origin of this deviation most likely lies in correlations
between $G_2(t,\tau)$ and $\langle I_p(t)\rangle_p$.
We are currently investigating the origin of this issue.

The measurement noise is expected to increase as the number of
pixels over which the intensity correlation is averaged,
$n_{pix}$, decreases. More precisely, the variance of the CCD
electronic (or dark) noise and read-out noise should scale as
$n_{pix}^{-1}$, because of the Central Limit Theorem. Similarly,
even in the absence of CCD noise, all pixel-averaged quantities in
Eq.~(\ref{eq:cI}) will fluctuate as the speckle pattern is renewed
over time, because $n_{pix}$ is finite \cite{Goodman1975}. By
invoking again the Central Limit Theorem, we expect the variance
of these fluctuations to exhibit the same $n_{pix}^{-1}$ scaling
as those due to the CCD noise. Therefore, the overall variance of
the measurement noise should decay with $n_{pix}$ as a power law
with exponent -1. The left panel of Fig.~\ref{VarVsnpix} shows
that indeed this behavior holds for all $\tau$, both smaller and
larger than the mean relaxation time $\tau_s$, thus further
supporting the hypothesis that the fluctuations of $c_I$ for
temporally homogeneous systems are due to the measurement noise.

The PDF and the variance of $c_I$ that we have so far introduced
are not sensitive to the detailed way $c_I$ fluctuates over time,
but only to its probability to assume a given value. To
investigate the time behavior of the fluctuations, we calculate
the time autocorrelation function of $c_I(t,\tau)$, defined as
\begin{equation}
\label{eq:autocI} {\rm Corr}(c_{I,\tau})(\Delta t) = \frac{\langle
(c_I(t,\tau) - \langle c_I \rangle_t)(c_I(t+\Delta t,\tau) -
\langle c_I \rangle_t)\rangle_t}{\langle (c_I(t,\tau) - \langle
c_I \rangle_t) ^2 \rangle_{t}}\,,
\end{equation}
where we have added the index $\tau$ to designate the time lag for
which $c_I$ is calculated. Note that with the above definition ${
\rm Corr}(c_{I,\tau})(0) = 1 $ and ${\rm Corr}(c_{I,\tau})(\Delta
t) = 0$ for uncorrelated signals (e.g. for $\Delta t \rightarrow
\infty $). The quantity introduced in Eq.~(\ref{eq:autocI}) is a
fourth-order time correlation function, since $c_I$ itself is a
time correlation function. In order to avoid any confusion, in the
following we shall refer to the argument $\Delta t$ of ${\rm
Corr}(c_{I,\tau})$ as to the $time$ $delay$, as opposed to the
$time$ $lag$ $\tau$ for $c_I$. In the right panel of
Fig.~\ref{VarVsnpix} ${\rm Corr}(c_{I,\tau})(\Delta t)$ is plotted
as a function of $\Delta t$ normalized by the relaxation time
$\tau_s$, for the same four time lags as in the previous figures
of this section. As it can be seen, ${\rm Corr}(c_{I,\tau})(\Delta
t)$ decays to 0 on a time scale comparable to the relaxation time
of the sample, $\tau_s$. Indeed, the data can be fit by an
exponential relaxation, whose characteristic time varies between
$0.21 \tau_s$ and $0.31 \tau_s$ (for clarity, only the fit for the
data collected at $\tau = 1.5$ sec is shown, as a solid line, in the
figure). At first sight this behavior may seem in contrast with
the assumption that here the fluctuations of $c_I$ are due to the
measurement noise, which is expected to be uncorrelated. Indeed,
this is the case for the CCD noise, as we tested by taking a
series of dark images. However, we recall that the fluctuations of
the quantities $\langle I_p(t)I_p(t+\tau)\rangle_p$ and $\langle
I_p\rangle_p$ used to compute $c_I$ are due not only to the CCD
noise, but also to the finite number of speckles that are sampled
when averaging over the CCD pixels. This latter contribution to
the noise of $c_I(t,\tau)$ is correlated over a time scale
comparable to $\tau_s$, for all time lags $\tau$. To demonstrate
it, let us consider as an example the fluctuations of
$c_I(t,\tau)$ for $\tau \gg \tau_s$, so that $\langle
c_I(t,\tau)\rangle_t = 0$. A measurement of $c_I$ at a given time
$t$ will yield a value $\epsilon$ that is in general slightly
different from zero, because $c_I$ is averaged over a finite
number of pixels. If the measurement is repeated at a later time
$t + \Delta t$, with $\Delta t \ll \tau_s$, the outcome will be
essentially the same, since the system configuration is almost
frozen on a time scale $\Delta t \ll \tau_s$ and thus the speckle
patterns recorded at time $t+ \Delta t$ and $t+\tau +\Delta t$ are almost
unchanged with respect to those recorded at time $t$ and $t+\tau$, respectively.
Therefore, $c_I(t+\Delta t,\tau)$ is highly correlated to
$c_I(t,\tau)$. As the delay is increased, the evolution of the
speckle images during $\Delta t$ will become increasingly relevant
and $c_I(t+\Delta t,\tau)$ and $c_I(t,\tau)$ will become less and
less correlated. Since the speckle fields are renewed completely
after a few $\tau_s$, one expects ${\rm Corr}(c_{I,\tau})(\Delta
t)$ to decay with a characteristic time close to $\tau_s$, as
observed in the right panel of Fig.~\ref{VarVsnpix} (note that
similar arguments can be invoked for all $\tau$).

\section{Statistical Analysis of TRC data for temporally heterogeneous dynamics}
\label{Hetero}

We now turn to TRC data collected for jammed systems and show that
the statistical properties of $c_I$ are very different from those
reported in the previous section for dilute Brownian particles.
Although we will focus in turn on different systems, such as a
micellar polycrystal, the colloidal paste cited above, and a
shaving cream foam, we stress that most of the features that will
be shown appear to be quite general. As we shall discuss in the
following, these differences indicate that the slow dynamics of
jammed systems is temporally heterogeneous.

  \begin{figure}
\centering{\resizebox{12.5cm}{!}{\includegraphics{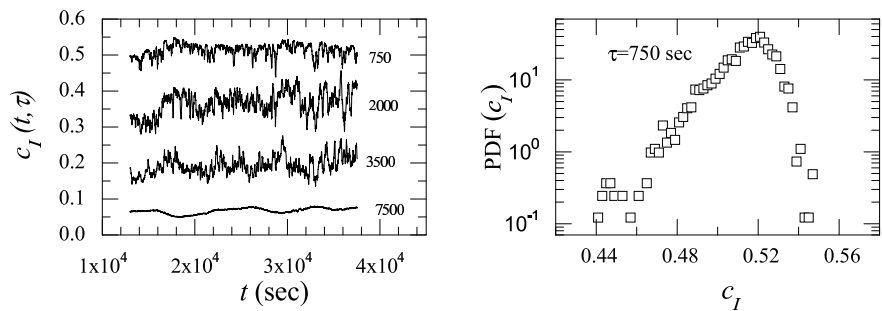}}}
\caption{ \label{cIF108}
For a micellar polycrystal are shown: Left panel: time variation
of $c_I(t,\tau)$ for several lags (the curves are labelled by the
normalized lag $\tau / \tau_s$). The relaxation time of $g_2-1$ is
$\tau_s = 3158.5$ sec. Right panel: for the same data, PDF of $c_I$
for $\tau / \tau_s$ = 0.237. Note the highly skewed shape of the
PDF, due to the marked drops of correlation visible in the left
panel. For the longest $\tau$, a Gaussian shape of the PDF is
recovered (not shown in the figure).}
   \end{figure}

Figure~\ref{cIF108} shows TRC data collected in the single
scattering geometry ($\theta = 45$ deg) for the slow dynamics of
the network of defects of a micellar polycrystal (more details on
the sample and its average dynamics can be found in
Ref.~\citenum{LucaFaraday2003} and the references therein). For
this system, the time-averaged intensity autocorrelation function
is a ``compressed" exponential: $g_2(\tau) - 1 = \beta \exp[-(\tau
/ \tau_s)^p]$, with $\tau_s = 3160$ sec and $p \approx 1.7$.
Similarly to the case of the colloidal paste data shown in the
upper panel of Fig.~\ref{cI2D}, the fluctuations of $c_I$ are very
different from those measured for a Brownian suspension (compare
to Fig.~\ref{cIpoly}, right panel). As discussed previously, the
large drops of $c_I(t,\tau)$ below its mean value indicate that
occasionally the sample undergoes a rearrangement that is much
larger than the typical change in configuration over a time
$\tau$. Therefore, the TRC data unambiguously show that the
dynamics of both the colloidal paste and the micellar polycrystal
is intermittent. The PDF of $c_I(t,\tau = 750~\rm{  sec})$ is shown
in the right panel. Because of the drops of the signal visible in
the left panel, the PDF severely deviates from the Gaussian shape
observed for the Brownian sample (compare to Fig.~\ref{PDFpoly})
and is strongly skewed to the left, the left tail being close to
an exponential decay. Interestingly, the shape of the PDF depends
on the time lag at which $c_I$ is measured. Without discussing the
details of this dependence, we remark that for $\tau \gg \tau_s$
the PDF recovers a Gaussian shape (data not shown). This behavior
can be understood by assuming that the average time between
rearrangements is much shorter than the mean relaxation time
$\tau_s$. Accordingly, a large number of rearrangements occur
between any two images taken at a lag $\tau \gg \tau_s$ and the
PDF is Gaussian because of the Central Limit Theorem. In other
words, on these time scales the dynamics is due to such a large
number of individual rearrangement events that it can be regarded
as continuous, much as for a Brownian suspension the dynamics is
continuous because it arises from the motion of a very large
number of particles, even though each individual particle moves in
a jerky fashion.

   \begin{figure}
\centering{\resizebox{12.5cm}{!}{\includegraphics{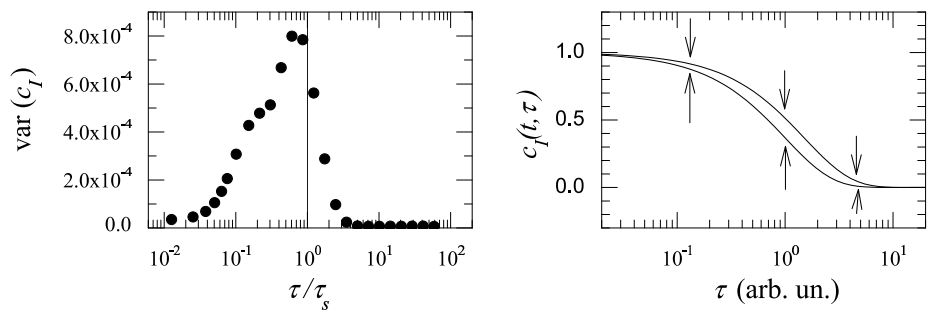}}}
\caption{ \label{VarVsTauPaste}
Left panel: $\tau$ dependence of the variance of $c_I$ for a dense
colloidal paste. Data are taken in the DWS transmission geometry.
Note the peak of ${\rm var}(c_I)$ for $\tau \approx \tau_s$ and
the large value of the fluctuations at the peak, compared to that
for $\tau \gg \tau_s$. Right panel: schematic representation of
the impact of a fluctuation of the relaxation time $\tau_s$ on the
value of $c_I$. As shown by the arrows, the change of $c_I$ at a
fixed lag is maximum for $\tau \approx \tau_s = 1$.}
   \end{figure}

We continue the comparison between the statistical properties of
TRC data for jammed systems and those for fluid systems by
contrasting the $\tau$ dependence of the variance of $c_I$ for the
colloidal paste introduced in Sect.~\ref{TRCSec} to that for the
diluted Brownian suspension, shown previously in
Fig.~\ref{VarVsTau}. As can be seen in Fig.~\ref{VarVsTauPaste},
for the colloidal paste ${\rm var}(c_I)$ is peaked around $\tau
\approx \tau_s$ (for this sample, $g_2(\tau) -1 \approx \beta
\exp(-\tau/ \tau_s)$, with $\tau_s = 1580$ sec), instead of
decaying with $\tau$, as for the Brownian particles. We stress
that the peaked shape is conserved even when plotting
parametrically ${\rm var}(c_I)$ $vs$ $(\langle c_I \rangle_t
+1)^2$ (data not shown), contrary to the case of the Brownian
suspension (see Fig.~\ref{VarVsTau}, right panel). Moreover, the
peak value of the variance is 2 orders of magnitude larger than
the measurement noise at large $\tau$, whereas for the Brownian
sample the difference between the smallest and largest value of
${\rm var}(c_I)$ is only about a factor of 3. Therefore, both the
presence of a peak and the larger amplitude of the fluctuations
clearly differentiate the dynamics of the pastes from that of the
Brownian particles, even though the shape of the time-averaged
correlation function is the same for both systems, i.e. an
exponential decay. The presence of a peak in ${\rm var}(c_I)$ can
be understood, at least qualitatively, by assuming that the
characteristic time of the relaxation of $c_I$ fluctuates with
time, $\tau_s = \tau_s(t)$, rather than being constant. To
illustrate schematically this point, we show in the right panel of
Fig.~\ref{VarVsTauPaste} two exponential decays with slightly
different relaxation times (in arbitrary units), $\tau_s = 1$ and
1.5, respectively. These two curves are meant to be representative
of the typical behavior of $c_I$ $vs$ $\tau$ measured at two
different times $t_1$ and $t_2$. As indicated by the arrows, the
resulting change in correlation at a given lag is maximum for
$\tau \approx \tau_s$ and decreases both at small and large lags,
in qualitative agreement with the curve shown in the left panel.
For our jammed systems, the physical origin of the fluctuations of
$\tau_s$ most likely lies in the fluctuation of the number, or the
size, of the regions that rearranged by each event. More work is
currently being carried on to explore quantitatively this concept.

   \begin{figure}
\centering{\resizebox{12.5cm}{!}{\includegraphics{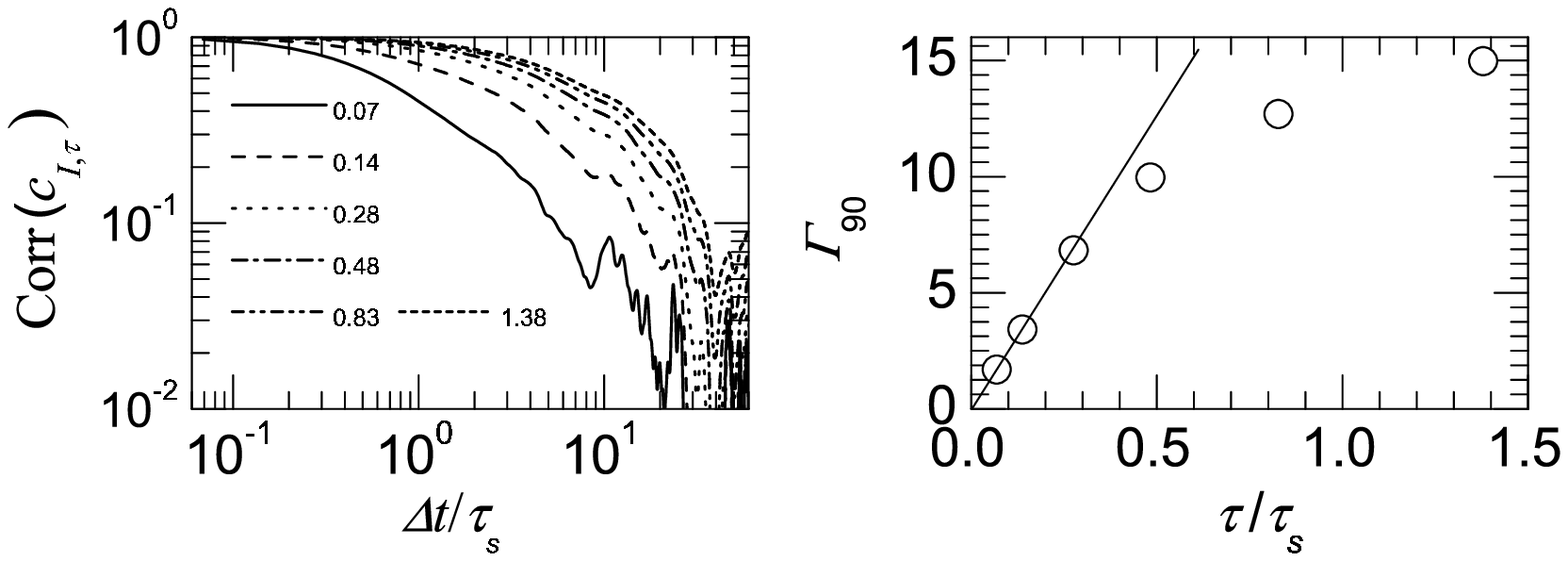}}}
\caption{ \label{CorrcIFoams}
For DWS measurements on a shaving cream foam are shown: Left
panel: the time autocorrelation of $c_I$, for various lags $\tau$,
as a function of the normalized delay $\Delta t / \tau_s$. The
curves are labelled by $\tau / \tau_s$. Contrary to the case of
Brownian particles, there is a strong $\tau$ dependence of the
decay rate of ${\rm Corr}(c_{I,\tau})(\Delta t)$. Right panel:
$\tau$ dependence of the relaxation time of the curves shown in
the left panel (symbols, see text for details). For the shortest
lags, $\Gamma_{90}$ is proportional to $\tau$, as indicated by the
straight line.}
   \end{figure}

To investigate how the fourth-order correlation function ${\rm
Corr}(c_{I,\tau})(\Delta t)$ behaves when the dynamics is
temporally heterogeneous, we analyze TRC data collected in the DWS
transmission geometry for a shaving cream foam. Previous work by
Durian $et~al.$ has shown that the dynamics of a foam is
intermittent, due to bubble rearrangements that are localized both
in time and in space \cite{DurianScience1991}. Therefore, the foam
represents an ideal system for testing the TRC method. In our
experiment, we find that the time-averaged correlation function
$g_2 - 1$ is well fitted by an exponential decay with
characteristic time $\tau_s = 0.29$ sec, in agreement with
Ref.~\citenum{DurianScience1991}. The left panel of
Fig.~\ref{CorrcIFoams} shows ${\rm Corr}(c_{I,\tau})$ vs. the
normalized delay $\Delta t / \tau_s$ for various lags $\tau$. In
sharp contrast with the behavior of the fourth-order correlation
for the Brownian sample (see Fig.~\ref{VarVsnpix}, right panel),
for the foam ${\rm Corr}(c_{I,\tau})$ strongly depends on the lag
at which $c_I$ is calculated, the decay time of ${\rm
Corr}(c_{I,\tau})$ increasing with $\tau$. To characterize this
dependence, for each $\tau$ we extract from the initial relaxation
of ${\rm Corr}(c_{I,\tau})$ a (normalized) characteristic time,
$\Gamma_{90}$, which we define as the (normalized) delay for which
${\rm Corr}(c_{I,\tau}) = C_{th}\equiv 0.90$ \footnote{Similar results are
obtained for different $c_{th}$, provided that $C_{th}>0.7$.}.
We plot $\Gamma_{90}$ as a function of
$\tau$ in the right panel of Fig.~\ref{CorrcIFoams}: as it can be
seen, for small lags $\Gamma_{90}$ is proportional to
$\tau$, whereas it deviates from this linear trend as $\tau$
approaches the relaxation time of $g_2-1$. Therefore, for the foam
the fluctuations of $c_I(t,\tau)$ are correlated over a time scale
proportional to $\tau$, at least as long as the time lag is not
too large.

   \begin{figure}
\centering{\resizebox{12.5cm}{!}{\includegraphics{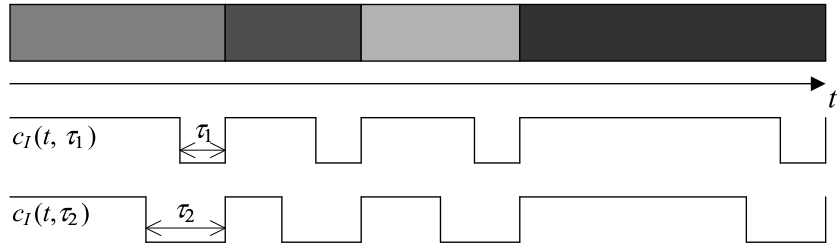}}}
\caption{ \label{ToyModel}
Schematic representation of the evolution of a system, such a
foam, whose dynamics is intermittent: each rectangle represents a
(slightly) different configuration of the system. The time
evolution of $c_I$ for two values of the lag, $\tau_1$ and
$\tau_2$, is shown below the time axis: $c_I$ drops whenever the
two speckle images that are correlated are taken for different
sample configurations. Note that the width of the dips is equal to
the time lag.}
   \end{figure}

In order to explain this behavior, we represent schematically the
evolution of a system whose dynamics is intermittent in
Fig.~\ref{ToyModel}. The rectangles at different grey levels
represent different configurations of the system: the evolution
occurs via discrete ``jumps" ---or rearrangements--- from a frozen
configuration to the next one, rather than through a continuous
change. The life time of the configurations is assumed to vary
randomly. The degree of correlation between pairs of speckle
images drops whenever the two images are taken for different
configurations, while it stays high within the same configuration
(see the sketch of $c_I$ below the time axis in
Fig.~\ref{ToyModel}, for two values of $\tau$). This implies that
the dips of $c_I$ have a width equal to the time lag $\tau$, and
that their raising edge is aligned with the change in
configuration, thus allowing one to identify precisely the time at
which the rearrangement event occurred. Accordingly, for any given
$\tau$, the TRC signal consist of a random sequence of downward
pulses of width $\tau$. The autocorrelation of such a signal
decays over a time scale of the order of the width of the pulses,
i.e. over a time of order $\tau$. Therefore, this rather crude
picture provides some explanation to the proportionality between
the decay time of ${\rm Corr}(c_{I,\tau})$ and $\tau$ that is
observed, at small $\tau$, for the foam. Note that as the lag
increases the likelihood of two or more rearrangements to occur
during a time $\tau$ becomes increasingly relevant. These multiple
rearrangements would alter the shape of the dips of $c_I$, thus
reducing the decay time of ${\rm Corr}(c_{I,\tau})$ with respect
to what is expected by extrapolating the small $\tau$ regime, as
observed in Fig.~\ref{CorrcIFoams} (right panel). Of course, a
more realistic description of the dynamics should take into
account the possibility that the rearrangement be more gradual and
the dynamics within a given configuration not completely frozen.
Moreover, in a macroscopic scattering volume many rearrangements
are likely to occur at any given time, so that in practice $\tau$
would be always longer than the mean life of the configuration.
This being the case, a more appropriate approach would be to
relate the fluctuations of $c_I(t,\tau)$ to those of the number of
rearrangements during a time $\tau$, rather to the occurrence of
single events. We are currently exploring this scenario.


\section{Conclusions}
\label{Comp}

We have shown that, contrary to traditional light scattering
techniques, TRC experiments have the capability of discriminating
between continuous and temporally heterogeneous dynamics. Among
the numerous ways one could analyze the TRC time series, three
methods seem particularly interesting: the characterization of the
PDF of the fluctuations of $c_I$, the study of the $\tau$
dependence of the width of the distributions, ${\rm
var}(c_{I,\tau})$, and the investigation of the temporal
correlations via ${\rm Corr}(c_{I,\tau})$. Other possible
approaches may include a wavelet analysis of $c_I$ and the
investigation of the fluctuations in the framework of chaos
theory. Quite generally, TRC measurements indicate that the slow
dynamics of soft jammed materials is temporally heterogeneous, in
striking analogy with recent numerical and experimental work on
hard condensed matter glasses (for a review, see
Refs.~\citenum{GlotzerNonCrystSolids2000,EdigerAnnuRevPhysChem2000}).
We conclude by a brief, and partial, overview of some of the
recent work related to the investigation presented in this paper,
either because of the techniques that are used or in view of the
results that are obtained. Dixon and Durian have introduced a
method termed Speckle Visibility Spectroscopy (SVS), where the
fast dynamics of a multiply scattering sample is studied by
measuring with a CCD the time-dependent contrast of the speckle
pattern, defined by $ \langle I_{p}(t)^2 \rangle_p / \langle
I_{p}(t) \rangle_{p}^2 - 1$ \cite{DixonPRL2003}. Note that, in our
notation, this quantity corresponds to $c_I(t,\tau = 0)$. If the
speckles fluctuate on a time scale shorter than or comparable to
the CCD exposure time, the image will be blurred and the contrast
reduced. Therefore, SVS provides information on the dynamics on a
time scale comparable to the exposure time (typically from a
fraction of msec to a fraction of sec), while in the TRC method
the shortest time scale is set by the CCD acquisition rate
(typically a few tens of images per second) and the longest one
can be as large as several hours. Other similarities exist between
the TRC and the work by Lemieux and Durian, who use fourth-order
intensity correlation functions to measure intermittent
dynamics\cite{LemieuxJOSA1999,LemieuxAPPOPT2001}. Although the
quantity ${\rm Corr}(c_{I,\tau})$ introduced here is also a
fourth-order intensity correlation function, we note that in
Refs.~\citenum{LemieuxJOSA1999,LemieuxAPPOPT2001} a point detector
is used and no pixel average is performed, as opposed to the way
$c_I$ is calculated. An extended time average is thus required to
reduce the measurement noise; contrary to TRC, no time resolution
can be obtained: it is possible to access some of the $average$
statistical properties of the fluctuations, not their detailed
temporal behavior.
An interesting parallel can be made between the work presented
here and the investigation of temporally heterogeneous dynamics in
2-D electron systems\cite{BogdanovichPRL2002,JaroszynskiPRL2002}.
Following a method originally designed by Weissman in the context
of spin glasses\cite{WeissmanRevModPhys1993}, a time series of the
fluctuations of the resistance (or the conductivity) is divided
into shorter segments. For each segment, the power spectrum of the
fluctuations is calculated. Finally, the so-called second spectrum
is obtained by calculating the power spectrum of the time series
of spectra calculated for the individual data segments. Because of
the relationship between power spectrum and autocorrelation
function of a signal, the fourth-order autocorrelation function
${\rm Corr}(c_{I,\tau})$ calculated here for TRC represents the
time-domain analogue of the analysis proposed in
Ref.~\citenum{WeissmanRevModPhys1993}.
The PDF of the fluctuations of the dynamics of a polymeric glass,
as obtained from dielectric measurements, has been reported by
Buisson $et~al.$\cite{BuissonJPCM2003,BuissonEurophysLett2003}.
Skewed, non-Gaussian PDFs are observed, similar to that reported
in Fig.~\ref{cIF108}. Non-Gaussian distributions are also observed
in simulations of spin glasses, where the dynamics was averaged
over subregions instead of over the whole simulation
box\cite{CastilloPRL2002,CastilloPRB2003}. Interestingly, for both
our data (Fig.~\ref{cIF108}) and the simulations on spin glasses,
the shape of the PDF is very close to the universal non-Gaussian
PDF proposed for the fluctuations in highly correlated systems
\cite{BramwellPRL2000}. In our case, however, we recall that the
shape of the PDF depends on the lag $\tau$ and that for $\tau \gg
\tau_s$ a Gaussian distribution is recovered.
Several recent works have focussed on the variance of the
fluctuations of the dynamics in glassy systems. In particular,
La\v{c}evi\'c {\it et al.}\cite{Lacevic2002} introduce in
simulations of glass formers a time-dependent ``order parameter''
$Q(t)$ that compares the system configuration at two times
separated by $t$ and then calculate the (normalized) variance of
$Q(t)$, $\chi_4(t)$. These quantities are analogous to our $c_I$
and ${\rm var}(c_{I})$, respectively, since $c_I$ measures the
degree of overlap between sample configurations separated by a lag
$\tau$. Indeed, $\chi_4(t)$ is found to be peaked around the
characteristic time of the final relaxation of $Q(t)$, much as, in
our experiments, ${\rm var}(c_{I})$ is peaked around $\tau_s$.
As a final remark, we point out that in most of the recent work on
heterogeneous dynamics in glassy systems, a key role is played by
spatial heterogeneity and spatial correlations between the
rearrangements. In our experiments, each CCD pixel is flooded by
light issued by the whole scattering volume (see
Fig.~\ref{apparatus}). Accordingly, any information on the spatial
localization of the rearrangements is lost and only global
temporal heterogeneity can be detected. We are currently working
on a modified TRC scheme to overcome this limitation and achieve
space- and time-resolved scattering measurements.

\section*{Acknowledgements}
A. D., P. B., and L. C. acknowledge CNES (grant no.
03/CNES/4800000123), the French Ministry of Research (ACI ``Jeunes
chercheuses, jeunes chercheurs'' JC2076) and the European
Community (grant MRT-CT-2003-504712) for supporting this work. H.
B. and V. T. were supported by the Swiss National Fond (grant
20-65019.01). The collaboration between Montpellier and Fribourg
groups is supported by CNRS (PICS no. 2410). It is a pleasure to
thank E. Pitard and L. Berthier for many useful discussions and D.
Popovic for bringing to our attention the work by M. B. Weissman.

\end{document}